\documentclass[letterpaper]{article} % DO NOT CHANGE THIS
\usepackage{aaai24}  % DO NOT CHANGE THIS
\usepackage{times}  % DO NOT CHANGE THIS
\usepackage{helvet}  % DO NOT CHANGE THIS
\usepackage{courier}  % DO NOT CHANGE THIS
\usepackage[hyphens]{url}  % DO NOT CHANGE THIS
\usepackage{graphicx} % DO NOT CHANGE THIS
\usepackage{natbib}  % DO NOT CHANGE THIS AND DO NOT ADD ANY OPTIONS TO IT
\usepackage{caption} % DO NOT CHANGE THIS AND DO NOT ADD ANY OPTIONS TO IT
\usepackage{algorithm}
\usepackage{algorithmic}

\usepackage{newfloat}
\usepackage{listings}
\DeclareCaptionStyle{ruled}{labelfont=normalfont,labelsep=colon,strut=off} % DO NOT CHANGE THIS
\floatstyle{ruled}
\newfloat{listing}{tb}{lst}{}
\floatname{listing}{Listing}
\title{Enhancing Cognitive Diagnosis using Un-interacted Exercises: A Collaboration-aware Mixed Sampling Approach}
\author{
    Haiping Ma\textsuperscript{\rm 1}, 
    Changqian Wang\textsuperscript{\rm 1}, 
    Hengshu Zhu\textsuperscript{\rm 2}, 
    Shangshang Yang\textsuperscript{\rm 3}\thanks{Corresponding Authors.}, \\
    Xiaoming Zhang\textsuperscript{\rm 1}, 
    Xingyi Zhang\textsuperscript{\rm 4\footnotemark[1]
}}

\affiliations{
    \textsuperscript{\rm 1}Department of Information Materials and Intelligent Sensing Laboratory of Anhui Province, \\Institutes of Physical Science and Information Technology, Anhui University, China\\
    \textsuperscript{\rm 2}Career Science Lab, BOSS Zhipin, China\\
   \textsuperscript{\rm 3} School of Artificial Intelligence, Anhui University, China\\
    \textsuperscript{\rm 4}School of Computer Science and Technology, Anhui University, China\\
    hpma@ahu.edu.cn, xmzhang@ustc.edu, \\
    \{changqian.wang.dl, zhuhengshu, yangshang0308, xyzhanghust\}@gmail.com
    
}

\usepackage{bibentry}
\usepackage{amsmath}
\usepackage{multirow}
\usepackage{tabularx}
\usepackage{multicol}
\usepackage{subcaption}
\usepackage{caption}
\usepackage{booktabs}
\usepackage{graphicx}
\usepackage{amssymb}
\usepackage{appendix}
\begin{document}

\maketitle

\begin{abstract}
Cognitive diagnosis is a crucial task in computational education, aimed at evaluating students' proficiency levels across various knowledge concepts through exercises. Current models, however, primarily rely on students' answered exercises, neglecting the complex and rich information contained in un-interacted exercises. While recent research has attempted to leverage the data within un-interacted exercises linked to interacted knowledge concepts, aiming to address the long-tail issue, these studies fail to fully explore the informative, un-interacted exercises related to broader knowledge concepts. This oversight results in diminished performance when these models are applied to comprehensive datasets. In response to this gap, we present the \textbf{C}ollaborative-aware \textbf{M}ixed \textbf{E}xercise \textbf{S}ampling (\textbf{CMES}) framework, which can effectively exploit the information present in un-interacted exercises linked to un-interacted knowledge concepts. Specifically, we introduce a novel universal sampling module where the training samples comprise not merely raw data slices, but enhanced samples generated by combining weight-enhanced attention mixture techniques. Given the necessity of real response labels in cognitive diagnosis, we also propose a ranking-based pseudo feedback module to regulate students' responses on generated exercises. The versatility of the CMES framework bolsters existing models and improves their adaptability. Finally, we demonstrate the effectiveness and interpretability of our framework through comprehensive experiments on real-world datasets.
\end{abstract}
\section{Introduction}
%Cognitive diagnosis models typically comprise students, exercises, and knowledge concepts. Their aim is to model students' characteristics by analyzing their historical interaction logs, thereby revealing students' knowledge states and cognitive levels.

%如图~\ref{introduction}所示，Existing cognitive diagnosis studies are based on the historical response logs between students and exercises, as well as the associations between exercises and knowledge concepts for modeling. They believe exercise interactions provide the greatest diagnostic value for students, while overlooking the information contained in un-interacted exercises. 实际上每个学生交互的习题数量只占了整个题库的很小一部分，而未交互的题目的信息是复杂且丰富的。

\begin{figure}[t]
    \centering
    
 \includegraphics[width=0.9\linewidth]{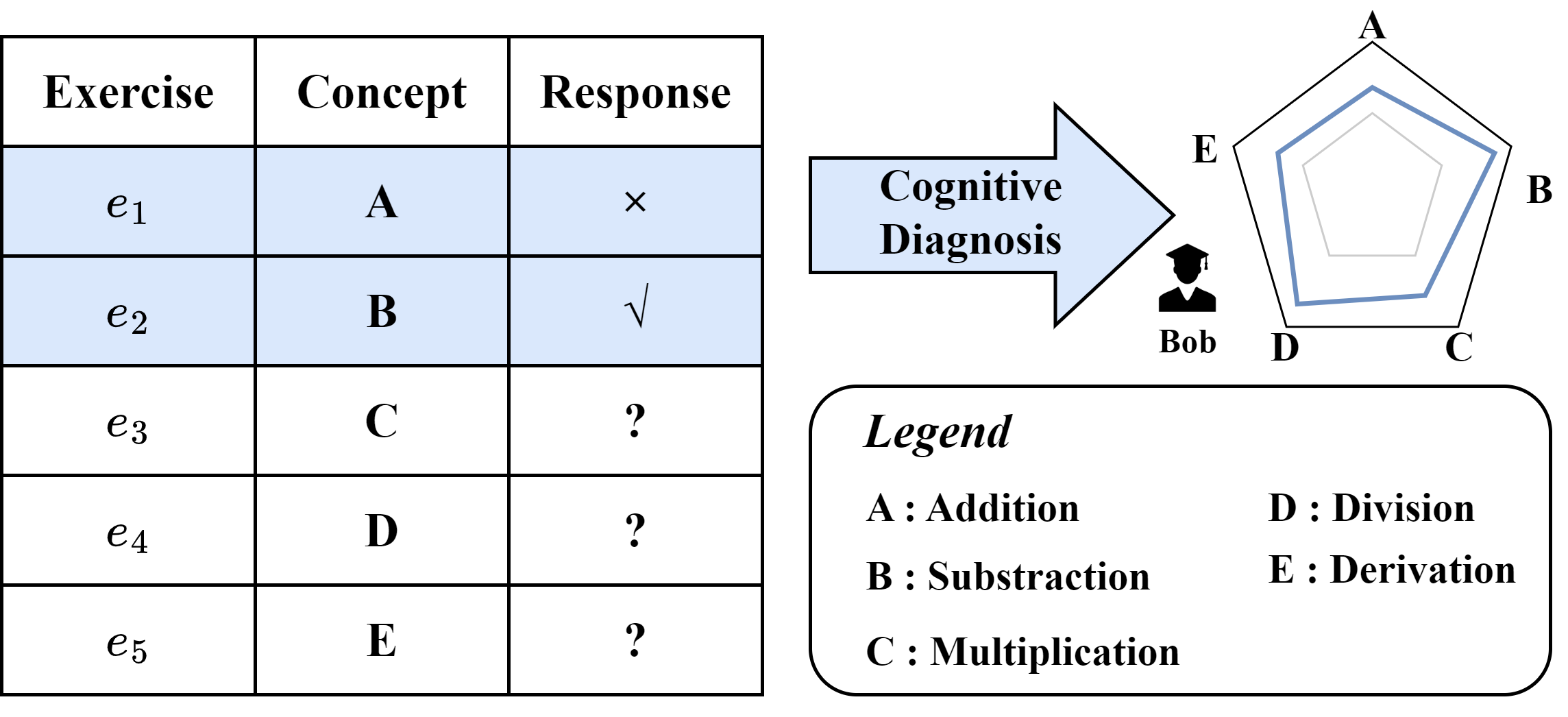}
     \caption{Illustration of cognitive diagnosis. Mainstream cognitive diagnosis models derive diagnosis results from students' response logs. $e_1$ and $e_2$ are the exercises that Bob has interacted with, while the remaining exercises are those that he has not interacted with.}
     \label{introduction}
\end{figure}

Amid the rapid advancement of computer-aided education, cognitive diagnosis has garnered increasing attention~\cite{applicationofirt,y1,q1}. As a crucial task in intelligent education, cognitive diagnosis aims at evaluating students’ proficiency levels across various knowledge concepts through exercises. As illustrated in Figure~\ref{introduction}, existing cognitive diagnosis studies are based on the historical response logs between students and exercises, as well as the associations between exercises and knowledge concepts for modeling. They believe exercise interactions provide the greatest diagnostic value for students, while overlooking the information contained in un-interacted exercises. In practice, each student's interaction with exercises represents a mere fraction of the complete exercise bank, with the un-interacted exercises containing intricate and extensive information.
%Existing cognitive diagnosis studies are based on the historical response logs between students and exercises, as well as the associations between exercises and knowledge concepts for modeling. They believe exercise interactions provide the greatest diagnostic value for students, while overlooking the information contained in un-interacted exercises. We devide 20\% of the full ASSISTments dataset~\cite{assist} as the test set. From the remaining data, we excerpted portions equivalent to 80\%, 70\%, 60\%, and 50\% of the full dataset as train sets, which are individually utilized to develop NeuralCD~\cite{ncd}, with model performance assessed on the shared 20\% test set. As illustrated in Figure \ref{introduction}, despite the well-established fact that training with less data leads to inferior model performance, from the standpoint of data interaction, compared to larger train sets, smaller train sets result in some interacted exercises being treated as un-interacted and neglected for model development. Thus, even at 80\% train set utilization, latent information abides in probable un-interacted data that could additionally augment cognitive diagnosis model performance.
% 现有的认知诊断研究是基于学生和习题的历史response日志，以及习题与知识概念之间的联系进行建模他们认为基于交互的习题信息可以为学生提供最大的诊断价值，而忽略了未交互题目所蕴含的信息。我们将完整的通用数据集ASSISTments的20%数据作为测试集划分出来。然后从剩余的数据中,我们分别提取出相对于完整数据集的80%、70%、60%和50%部分作为训练集。我们将得到的基于完整数据集的80%、70%、60%和50%的4个训练集分别训练模型然后在共同的20%的测试集评估模型的性能。如图\ref{introduction}所示，尽管众所周知，用较少的数据训练模型会得到较差的性能，但从数据交互的角度来看，这是因为较少的训练集相较于较多的训练集，将部分交互过的题目作为未交互习题而没有用来训练模型，因此尽管对于80%的训练集，依然有潜在的未交互数据蕴含有助于提升认知诊断模型性能的丰富信息。

In this paper, we attempt to leverage these un-interacted exercise information. One challenge with leveraging such un-interacted exercise information is the absence of students' potential response labels. Recent work in EIRS~\cite{eirs} makes the assumption that students will perform comparably on exercises related to the same knowledge concepts. EIRS aims to mitigate the long-tail problem (where insufficient interaction data results in skewed distributions) through similarity-oriented sampling of exercises associated with previously interacted concepts. Since the attained sample exercises convey analogous information to the interacted ones, constraining the acquired knowledge, this circumstance culminates in the method being unable to realize optimal performance on full datasets.
%对于利用这类未交互习题的信息的一个挑战是缺少学生的潜在作答标签。最近的一份工作EIRS suppose that a student performs similarly on exercises that share the same knowledge concept，它通过基于相似度的采样方式来利用交互过的知识概念关联的习题解决长尾问题(即过少的交互信息导致数据分布不均衡)。但由于采样到的题目与交互过的题目蕴含的信息量是相似的，导致采样得到的信息量有限，因此针对全量数据集并不能达到很好的效果。

Consequently, determining how to extract additional informative un-interacted exercises constitutes another challenge. Within the domain of recommender systems, informative negative samples are frequently utilized to train the system and enhance recommendation performance~\cite{pairwise1}. Accordingly, substantial research has investigated techniques for sampling informative negative instances~\cite{bpr,reinforcedNS,bayesianns}. However, owing to the distinctive nature of cognitive diagnosis models, which encompass intricate interrelationships among students, exercises, and knowledge concepts, sampling approaches from recommender systems are not transferable to cognitive diagnosis models.
%因此如何挖掘信息量多的未交互习题是另一个挑战。在推荐系统领域中通常利用信息量丰富的负样本来训练系统，以提升推荐性能，因此大量工作对如何采样信息量丰富的负样本进行了探索，但由于认知诊断模型的特殊性，其中涉及了学生与习题和习题与知识概念的复杂关系，使得推荐系统中的采样方法并不能适用于认知诊断模型。

To address the aforementioned challenges, we propose a general framework, namely Collaborative-aware Mixed Exercise Sampling (CMES) for cognitive diagnosis models. CMES extracts more informative diagnoses from the pool of un-interacted exercises and obtains students' potential response labels. Specifically, to improve the quality and efficiency of sampling, we preclude sampling exercises affiliated with students' interacted concepts, as well as exercises with potentially similar information to interacted ones. We cluster students founded on their response capabilities and collaborations. Throughout the sampling progression, we sample from other students' interacted exercise sets in different clusters. As interacted exercises encompass robust diagnostic intimations while un-interacted ones retain prospective heterogeneous information, we use mixing techniques to mix the sampled exercise information by injecting interacted exercise information into sampled exercises, obtaining more informative mixed samples. Finally, we design a ranking based pseudo feedback module to predict potential response situations for the sampled exercises, which is combined with the cognitive diagnosis task for joint learning.
%为了解决上述挑战，我们提出了适用于认知诊断模型的通用框架，Collaborative-aware Mixed Exercise Sampling (CMES)，从未交互的习题集中获取更丰富的信息，并得到学生对其潜在的回答标签。具体来说，为了提高采样质量与效率，我们不仅放弃采样学生已交互过的知识概念关联的习题，我们还不采样与已交互习题可能具有相似信息量的习题。我们根据学生的答题能力和学生之间的协同关系对学生进行聚类，在采样时从其他簇的学生交互过的习题集中进行采样。由于交互过的习题有强的诊断信息，而未交互的习题蕴含着潜在的多样化信息量，因此我们使用混合技术对采样到的习题的信息进行融合，将交互过的习题与采样到的习题的信息相互注入到彼此的信息中，以得到信息量更丰富的样本。最后，我们设计了一个基于排序学习的Pseudo Feedback Module对采样到的题目进行预测潜在的作答情况。

Our main contributions are summarized as follows:

\begin{itemize}
\item To fully leverage the latent information in un-interacted exercises for student diagnosis, we propose a generic sampling framework CMES for enhancing cognitive diagnosis tasks.
\item We specially design a learning-based pseudo feedback module that defines a learning-to-rank task assisting in the training of the cognitive diagnosis task,
% we design a Pseudo Feedback Module based on learning-to-rank that ranks students' potential cognitive abilities on mixed samples and interacted exercises using real response labels. This constrains students' potential answering abilities to derive reasonable pseudo labels via ranking.
\item We have conducted extensive experiments on real-world datasets to validate the effectiveness and scalability of our approach.
\end{itemize}
\section{Related Work}
In this section, we review the related work about cognitive diagnosis models and sampling strategies.

\subsection{Cognitive Diagnosis}
Cognitive diagnosis, a fundamental and critical task in education, aims to infer students' mastery of knowledge concepts. The early models IRT~\cite{irt} and DINA~\cite{dina} are two classic cognitive diagnosis models. Unlike IRT which hypothesizes unidimensional independence and adopts continuous latent variables to evaluate examinees' potential abilities, DINA is based on the attribute independence assumption and uses $0/1$ binary vectors to represent students' mastery of each attribute. MIRT~\cite{mirt}, as an extension of IRT, discards the unidimensionality and proposes that student proficiency is multidimensional, thus utilizing multiple latent traits to characterize students more comprehensively. NCD~\cite{ncd} firstly introduces neural networks into cognitive diagnosis, so as to capture the sophisticated student-exercise relationships. Afterwards, more neural network based approaches~\cite{y2,y3,y4,y5} are proposed, such as ECD~\cite{ecd} incorporates contextual features to facilitate more precise diagnosis of students' cognitive status. RCD~\cite{rcd} and SCD~\cite{scd} attempt to explore student-exercise-concept associations via graphs and conduct more delicate modeling of the interactions. Recent work of ICD~\cite{icd} further investigates the intrinsic correlations among knowledge concepts and quantitative relationships between exercises and concepts. Despite the remarkable progress, existing methods exclusively take advantage of interacted responses while overlooking the un-interacted yet more informative exercises.
%认知诊断是教育领域中的一项基本且重要的任务。早期的IRT和DINA是两种经典的认知诊断模型，DINA基于属性独立性假设，使用0/1二元向量表示学生对每个属性的掌握情况，与之不同的IRT是基于单维独立性假设，将学生能力和习题难度定义为一维连续的潜在变量来评估被试者的潜在能力。MIRT是IRT的一种改进模型，它放弃了IRT的单维度假设,而是假设学生能力是多维的,使用多个潜在变量更全面的刻画学生的特征。NCD首次将神经网络引入认知诊断模型，相较于传统的手动设计学生与习题的交互关系，NCD可以更加准确地捕捉到学生和习题之间的复杂关系。ECD引入了上下文特征等额外概念，帮助更准确地诊断学生的认知状态。RCD首次使用图来探索学生-习题-知识概念的内部关系，将交互关系进行了更加细致的建模。最近的工作ICD进一步探索了知识概念内部的交互关系和习题和概念之间的定量关系。尽管现有的方法都取得了很大的进展，但它们都是使用已有的交互信息，而忽略了信息更加复杂且丰富的未交互信息。
\subsection{Sampling Strategy}
Sampling strategies are extensively utilized in recommender systems, where sampling informative non-interacted instances close to positive samples facilitates models to better learn the boundary between positive and negative samples. Conventional recommender systems often adopt random negative sampling~(RNS)~\cite{bpr} and static popularity-based~ negative sampling~(PNS)~\cite{pns,pns1}, through which the attained negative samples are typically of low quality and fail to train models effectively. Dynamic negative sampling~(DNS)~\cite{dns} is an adaptive negative sampling approach, scoring each sample and using high-scored ones as negative samples for model training. Currently, GAN-based negative sampling~\cite{irgan,reinforcedns1,ipgan} prevails in recommender systems. Despite explorations into GANs, existing GAN-based sampling strategies often suffer from poor interpretability and inferior performance due to training instability. A graph data augmentation based negative sampling~\cite{mixgcf} augments the positive samples with negative sample information to delude the recommender and enhance its ability to distinguish the boundary. Due to the sophisticated student-exercise interactions that not only reply on answering records but also associations between exercises and concepts, transplanting negative sampling strategies from recommender systems into cognitive diagnosis faces challenges. Although previous work EIRS~\cite{eirs} has introduced sampling strategies into cognitive diagnosis, it essentially performs similarity-based sampling, where the attained samples carry comparable information to interacted ones and fail to provide extra diagnostic values. Inspired by the high-quality negative samples achieved in recommender systems, we propose a novel sampling strategy to obtain informative samples.

%采样策略在推荐系统中广泛使用，采样与正样本接近的信息丰富的未交互样本，可以使模型更好地学习正样本和负样本之间的边界。早期推荐系统中通常使用随机采样和基于流行度的静态负采样算法，然而通过这种方式获得的负样本通常样本质量低，无法有效地训练模型。DNS是一种自适应负采样算法，它对每个样本进行评分，使用高分样本作为负样本训练模型。当前基于GAN的负采样策略在推荐系统中流行使用，尽管现有工作对GAN进行了探索，但基于GAN的采样策略通常可解释性差，且因训练不稳定而导致模型性能差。一种基于图的数据增强的负采样策略是将负样本信息注入到正样本中，以此迷惑推荐系统，提高推荐系统学习正负样本之间边界的能力。由于学生对习题的交互信息更复杂，不仅依赖于答题情况，更依赖于习题与知识概念之间的联系，因此将推荐系统中的负采样策略应用到认知诊断模型中具有挑战性。尽管现有工作EIRS已经将采样策略应用在认知诊断模型中，但是实际上是一种基于相似度的采样方式，所获得的样本的信息与正样本所携带的信息量相当，无法提供更多的诊断信息。借鉴于推荐系统中负采样策略获得高质量负样本的思想，我们设计一种新的采样策略来得到信息量丰富的样本。
\section{Problem Statement}

For cognitive diagnosis, we define three entity groups: the student set $S = \{s_1, s_2, ..., s_N\}$ of size $N$; the exercise set $E = \{e_1, e_2, ..., e_M\}$ of size $M$; and the knowledge concept set $K = \{k_1, k_2, ..., k_C\}$ of size $C$. The exercise-concept relationship is defined by matrix $Q\in \mathbb{R}^{M\times C}$, where $Q_{i,j}=1$ if exercise $e_i$ involves concept $k_j$, else $Q_{i,j}=0$. We also define interaction logs as triplets $(s_i, e_j, r_{ij}) \in R$, where $e_j$ is called an interacted exercise of student $s_i$, $r_{ij} = 1$ if student $s_i$ correctly answered exercise $e_j$, else $r_{ij} = 0$ and $R$ is the interaction set. The un-interacted exercise set for student $s_i$ is defined by $U_{i} = E \backslash  E_{i}$, where $E_{i}$ is the interacted exercise set of student $s_i$. The knowledge concepts associated with the interacted exercises by student $s_i$ are called interacted knowledge concepts $K_i$, where $K_i\subset{K}$.

PROBLEM DEFINITION. \textit{Given student entity, exercise entity, knowledge concept entity, students' exercising response logs, the un-interacted exercises set $U_{i}$ for each student, and the exercise-knowledge relational matrix. Our goal is to leverage the un-interacted exercises to enhance the performance of cognitive diagnosis.}

\section{The Proposed CMES Framework}

In this section, we first briefly introduce the proposed framework, then elaborates on each module, and finally discusses how to train the cognitive diagnosis model with the proposed CMES framework. 
% This paper contains many symbols. For convenience, we have compiled all the symbols and their meanings in the appendix.

\begin{figure*}[htp]
 \centering
 \includegraphics[width=1.0\textwidth]{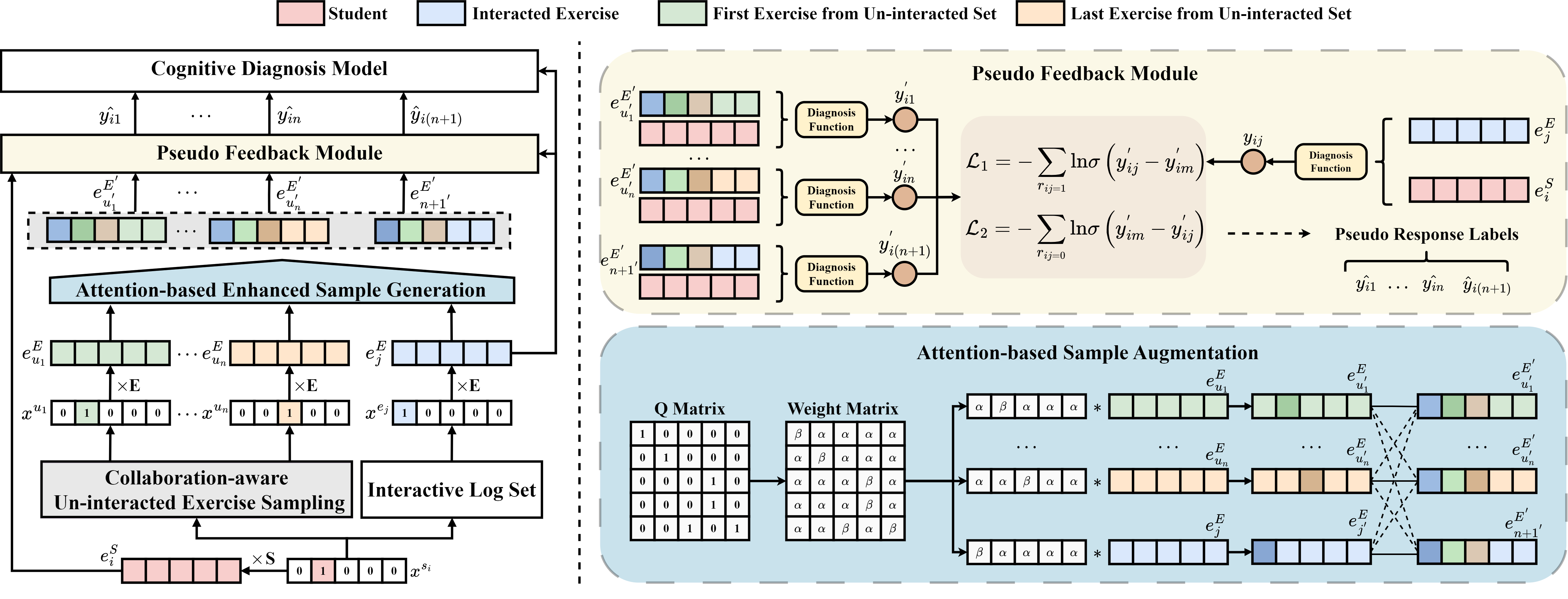}
\caption{The overview architecture of CMES: (a) The Sample Augmentation Module, which consists of Collaboration-aware Un-interacted Exercise Sampling and Attention-based Sample Augmentation; (b) The Pseudo Feedback Module; (c) Model Training via the Cognitive Diagnosis Task. }
 \label{framework}
\end{figure*}

\textbf{Overview.} The brief idea of this paper is to enhance cognitive diagnosis by sample augmentation using un-interacted exercises. For this aim, as shown in Figure \ref{framework}, our CMES framework comprises three key components: the sample augmentation module, the pseudo feedback module, and the extensible diagnosis module. The initial two modules aim to sample and blend informative exercises from the pool of un-interacted exercises for individual students, simultaneously evaluating potential feedback labels for the mixed exercises. More precisely, within the sample augmentation module, we group students based on their response capabilities to mitigate interference from exercises with limited information. We then proceed to sample and mix exercise information from other clusters. Once we acquire informative samples, the pseudo feedback module leverages interacted information to deduce students' feedback, subsequently generating pseudo response labels for each mixed sample. The final module~(i.e., the cognitive diagnosis module) employs the mixed exercises with pseudo labels and interaction records to deduce students' cognitive levels. Notably, our framework exhibits remarkable extensibility, seamlessly integrating supplementary data into existing methods, thereby enhancing their performance.
%我们的CMES框架不仅可以采样信息量丰富的习题，而且可以评估学生对采样习题的潜在作答表现。如图~\ref{framework}所示，CMES包括三个主要部分：the sample augmentation module,the pseudo feedback module，and the 可扩展诊断模块。具体来说，在sample augmentation module，我们通过学生的答题能力对学生聚类，以减少信息量有限的题目的干扰，然后从其他簇中采样并融合习题的信息。在得到信息丰富的样本后，pseudo feedback module利用已交互信息来推测学生的反馈，为每个样本生成伪反应标签。诊断模块利用采样到的样本和伪反应标签来推测学生的认知水平。特别是，我们的框架具有很好的可扩展性，因为它可以自然地将额外的数据混合到当前方法中并提高它们的性能。

\subsection{Sample Augmentation Module}

To thoroughly augment the information encompassed within the samples for each student, we first sample more informative exercises for each student from un-interacted exercises by clustering students. Subsequently, the sampled exercises are combined with the interacted exercises to generate novel samples, facilitated by attention mechanisms.

\subsubsection{Collaboration-aware Un-interacted Exercise Sampling}

This section focuses on the objective of selecting a specific number of exercises for each student $s_i$ from the un-interacted exercise set $U_{i}$. We posit the existence of two types of exercises within $U_{i}$ that offer limited supplementary information aimed at improving the accuracy of students' proficiency diagnosis. The initial type encompasses exercises related to the knowledge concepts within $K_i$ that have been interacted with by student $s_i$. This is facilitated by the understanding that student $s_i$'s proficiency with respect to the knowledge concepts in $K_i$ can be discerned from the interaction records. The second type consists of exercises accomplished by students exhibiting similar proficiency levels, drawing inspiration from the notion of collaboration. Given that these details can be somewhat captured by prevailing cognitive diagnosis models.

Thus, we structure the sampling process in the following manner. Initially, we partition students into $W$ groups based on their performance in exercises and the exercise-concept relational matrix $Q$. Subsequently, for the student $s_i$ with an interaction set $R_i$ of size $t$, we give preference to exercises that are commonly completed by peers within the remaining $W-1$ clusters, as these exercises have garnered more feedback. In other words, for student $s_i$, we draw a sample of $2n$ exercises, forming the candidate set $U_i^{cand} = \left \{u_1, u_2, ..., u_{2n}\right \} \subseteq U_{i}$ which intentionally excludes exercises linked to the knowledge concepts in $K_i$, and the value of $n$ serves as a hyperparameter.

\subsubsection{Attention-based Sample Augmentation}
Using the sampled $2n$ exercises ${U}_i^{cand}$ for student $s_i$, we combine these exercises with the interacted exercises $E_i$ to create a newly generated sample set, thereby augmenting our samples. More precisely, for each interacted exercise $e_j \in E_i$, we randomly select $n$ exercises~(denoted $U_{i,j}^{E}$) from the set ${U}_i^{cand}$. Subsequently, we combine these $n+1$ exercises, leveraging an attention mechanism to produce $n+1$ new samples. Consequently, for student $s_i$ who possesses an interacted exercise set $E_i$ containing $t$ exercises, we will generate a total of $t\times(n+1)$ new samples. Interacted exercises consistently provide substantial information, whereas sampled un-interacted exercises offer a range of diverse and informative insights. This mixing operation serves to balance the informativeness and diversity of samples, thereby enhancing the robustness and precision of student $s_i$'s diagnosis.

As we apply mixture to the vector representations of exercise instances, we initiate the embedding process of exercises by performing a matrix multiplication. Specifically, the one-hot vector $x^{e_{j}}$ for each exercise $e_j$, along with $x^{u_m}$ for exercise in $U_{i,j}^{E}$, is multiplied by a trainable matrix $E\in \mathbb{R}^{M\times d}$ to attain their initialized embedding representation $e_{j}^{E},e_{u_m}^E\in \mathbb{R}^{1\times d}$, where $M$ is the number of exercises, $d$ is the embedding size:
\begin{equation}
\small
e_{j}^{E}=x^{e_{j}}\times E,\quad e_{u_{m} }^{E}=x^{u_{m}}\times E.
 \end{equation}

Cognitive diagnosis models commonly utilize $e_{j}^{E}$ as the exercise $e_j$'s feature vector. The dimensions of $e_{j}^{E}$ correspond to the quantity of knowledge concepts, with each dimension representing an exercise attribute concerning the relevant concept. In this study, for a deeper understanding of exercises related to un-interacted knowledge concepts, it is essential to amplify the weights of correlated knowledge concepts in the information mixing process. As such, we propose to construct a weight matrix based on the $Q$ matrix denoted as $Q'\in \mathbb{R} ^{\mathrm {M\times C}}$, where $M$ and $C$ represent the number of exercises and knowledge concepts respectively:
\begin{equation}
\small
Q_{m,c}'=\left\{
\begin{aligned}
     \alpha,\ & \text{ if } Q_{m,c}=0\\
     \beta,\ & \text{ if } Q_{m,c}=1\\
\end{aligned}
\right.,
\end{equation}
where $Q_{m,c}$ denotes exercise $e_m$ is affiliated with knowledge concept $k_c$, $\alpha$ and $\beta$ represent hyperparameters, subject to $\alpha < \beta$. Then, we multiply the knowledge concept weight vector $Q_{j}'$ with the initialized embedding vector $e_{j}^{E}$ of exercise $e_j$ to obtain the exercise embedding vector $e_{j'}^{E}\in \mathbb{R}^{(n+1)\times d}$ with knowledge concept weight enhancement as follows:
\begin{equation}
e_{j'}^{E}=e_j^{E}\cdot Q_{j}'.
\end{equation}

Based on the embedding representation vectors of exercises, for each interacted exercise $e_j \in E_i$ and the randomly selected $n$ exercises $U_{i,j}^{E}$, we employ a self-attention network to mix them and obtain $n+1$ new embedding vectors. The embedding vectors of generated samples incorporate the learned target embedding as well as the information from one another. Here we adopt the\textit{ Scaled Dot-Product Attention} to capture the information among the sampled instances and interacted exercises: 
\begin{equation}
\small
\begin{aligned}
Q, K, V = &e_{j'}^{E}\times  W_Q,\  e_{j'}^{E} \times  W_K, \  e_{j'}^{E} \times  W_V\\
   &A_j= softmax ( \frac{Q\times K^T}{\sqrt{d} }  ) V
\end{aligned},
\end{equation}
where $W_Q\in \mathbb{R}^{d\times d}$, $W_K\in \mathbb{R}^{d\times d}$, and $W_V\in \mathbb{R}^{d\times d}$ are three trainable matrices. 
$A_j\in \mathbb{R} ^{ ( n+1  ) \times d}$ is the result computed by the attention module, 
representing a weighted vector that captures information from other exercises.
\iffalse
where $W_Q \in \mathbb{R} ^{\left ( n+1 \right ) \times d}$ contains $n+1$ exercises representations of the queries, $W_K \in \mathbb{R} ^{\left ( n+1 \right ) \times d}$ is the key matrix, $W_V \in \mathbb{R} ^{\left ( n+1 \right ) \times d}$ is the value matrix of the exercise attended, each row of the matrix $A \in \mathbb{R} ^{\left ( n+1 \right ) \times d}$ is added to $U_{i,j}^{E'}$, representing a weighted vector that captures information from other exercises. 
\fi
Then $A_j$ is taken as the $j$-th item~(i.e, $U_{i,j}^{E'}$) of the diagnosis-generated sample set  $U_{i}^{E'}= \{ U_{i,1}^{E'},U_{i,2}^{E'},\dots,U_{i,j}^{E'},\dots,U_{i,t}^{E'}  \}$, which encompasses $t \times \left ( n+1 \right ) $ samples for student $s_i$.

\subsection{Pseudo Feedback Module}
The samples generated by the sample augmentation module lack genuine response labels, which is necessary for cognitive diagnosis models. Therefore, within this module, we introduce a learning-to-rank task~\cite{pairwise} to deduce the corresponding pseudo response label for these generated samples, relying on the following assumption.

\textbf{Assumption.} We assume that the correct probability of student $s_{i}$ answering the interacted sample $e_{j}^{E}$ is greater than or equal to that of this student answering each generated sample, when $r_{ij}=1$. Otherwise, when $r_{ij}=0$, the ranking relationship is the opposite. This assumption is formally defined as follows:
\begin{equation}
\begin{aligned}
P\left ( r_{im} \right ) \le P\left ( r_{ij} \right ) \le 1,r_{ij} = 1,\\
P\left ( r_{im} \right ) \ge P\left ( r_{ij} \right ) \ge 0,r_{ij} = 0,
\end{aligned}
\end{equation}
where $r_{ij}$ is the real response label of student $s_i$ on the exercise $e_j$, $r_{ij}=1$ if student $s_i$ correctly answered exercise $e_j$, otherwise $r_{ij}=0$; $P\left ( r_{im} \right )$ and $P\left ( r_{ij} \right )$ denote the probabilities of student $s_{i}$ correctly answering the exercises $e_{u_{m}^{'}}^{E'}\in U_{i,j}^{E'}$ and $e_{j}^{E}$, respectively. 

Based on the above assumption, we simply define a learning objective, which is to maximize the following function:
\begin{equation}
\small
  \prod_{e_{u_{m}^{'}}^{E'} \in U_{i,j}^{E'}} 
 r_{ij}\times P_{i} (e_{j}^{E}> e_{u_{m}^{'}}^{E'} ) +
(1-r_{ij})  P_{i} ( e_{u_{m}^{'}}^{E'} > e_{j}^{E} ) ,
\end{equation}
where $P_{i}\left ( a> b \right )$ represents the probability of student $s_{i}$ correctly answering exercise $a$ is higher than that of correctly answering exercise $b$. 

We employ the BPR (Bayesian Personalized Ranking)~\cite{bpr} loss function to simplify the learning of the objective:
\begin{equation}
\small 
\begin{aligned}
    \mathcal{L}_{Feedback} =- \sum_{e_{u_{m}'}^{E'} \in U_{i}^{E'}}  ( \ r_{ij}*\ln{ \sigma  ( y_{ij}' -y_{im}'  )}  \\
    + (1-r_{ij})*\ln{\sigma  ( y_{im}' -y_{ij}' )} \ )  \\
\end{aligned},
\end{equation} 
where $\sigma \left ( \cdot \right ) $ is the sigmoid function, $y_{ij}'$ and $y_{im}'$ are obtained by the diagnosis function~(denoted by $f_1$) of the cognitive diagnosis model. Then, for each un-interacted exercise $e_{u_{m}'}$, we map $y_{im}'$ into $\hat{y_{im} } \in \left \{ 0,1 \right \} $ as the pseudo feedback response label of student $s_i$ on exercise $e_{u_{m}'}$, where $1$ indicates student $s_{i}$ may correctly answer the exercise $e_{u_{m}'}$, while $0$ indicates possible wrong answers. 

 % Ultimately, we obtained the label set $\hat{y}_{i,j}=\left \{ \hat{y}_{i1},\hat{y}_{i2},...,\hat{y}_{i(n+1)}\right \} \in \hat{y}_{i}$ corresponding to $U_{i,j}^{E'}$, where $\hat{y}_{i}$ signifies the set of labels that corresponds to $U_{i}^{E'}$.

\subsection{Cognitive Diagnosis Module}
\textbf{Learning Model with CMES Framework. }Our framework is applicable to any prevailing cognitive diagnosis model. The Sample Augmentation Module and Pseudo Feedback Module cater to student $s_i$ in the cognitive diagnosis model by providing personalized pairs of sampled exercises $U_i^{E'}$ and their corresponding pseudo feedback response labels. 

\textbf{Training.} Through predicting students' proficiency levels, we derive the ultimate mastery status for each student. In addition to employing the generated sample set $U_i^{E'}$, we also leverage the interaction set $R$ for diagnostic purposes. The loss function is defined as a composite of two components. Initially, we employ the frequently employed cross-entropy loss function~\cite{y11,y14}  within conventional CD models on the data from the interaction set $R$:
\begin{equation}
\small
\begin{aligned}
\mathcal{L}^{inter} = - \sum_{(s_{i},e_{j},r_{ij})\in \mathcal{R}}  & \left( r_{ij}\log y_{ij} \right)+ \left( 1 - r_{ij} \right) \log\left( 1 - y_{ij} \right),
\end{aligned}
\end{equation}
where $y_{ij}$ represents the proficiency prediction for student $s_{i}$ on exercise $e_j$ attained through diagnosis function of the cognitive diagnosis model~(denoted by $f_2$).

Then, we design a loss function for the mixed exercises:
\begin{equation}
\small
\begin{aligned}
\mathcal{L}^{un-inter} = - \sum_{s_{i}\in S}\frac{1}{\left| U_{i}^{E'} \right|} \sum_{e_{u_{m}'}^{E'} \in U_{i}^{E'}} \bigg( \left(\hat{y}_{im}\log y_{im}\right) & \\
 + \left(1-\hat{y}_{im}\right) \log\left(1-y_{im}\right) \biggr),
\end{aligned}
\end{equation}
where $y_{im}$ represents the proficiency prediction for student $s_{i}$ on mixed sample $e_{u_{m}^{'}}^{E'}$ attained through the cognitive diagnosis function $f_2$, $\hat{y}_{im}$ indicates the pseudo feedback label of student $s_i$ on $e_{u_{m}^{'}}^{E'}$. We optimize the cognitive diagnosis module using the following loss function:
\begin{equation}
\mathcal{L}_{CD}= \mathcal{L} ^{inter} + \mathcal{L} ^{un-inter}. 
\end{equation}

We optimize the entire framework using the following loss function:
\begin{equation}
\mathcal{L}_{CMES}= \mathcal{L}_{CD}\left ( \Theta _1 \right )  + \alpha \cdot\mathcal{L}_{Feedback}\left ( \Theta _2 \right ) , 
\end{equation}
where $\alpha$ is a balancing hyper parameter that weighs the two loss functions, $\Theta_1$ and $\Theta_2$ represent the training parameters of the Pseudo Feedback Module and the Cognitive Diagnosis Module respectively. It is worth noting that the diagnosis functions $f_1$ and $f_2$ in the Pseudo Feedback Module and the Cognitive Diagnosis Module apply the same cognitive diagnosis model but different parameters.

\section{Experiments}
As the key contribution of this work is to extend existing cognitive diagnosis models~(CDMs) to adaptively utilize un-interacted data, we compare the original CDMs and our optimized CDMs with the CMES framework (denoted as Orginal-CDMs and CMES-CDMs respectively) on real-world datasets to address the following research questions:
\begin{itemize}
\item \textbf{RQ1:} Can CMES-CDMs outperform Original-CDMs in terms of performance?
\item \textbf{RQ2:} How does our sample augmentation strategy outperform random sampling?
\item \textbf{RQ3:} Whether the performance of CMES is sensitive to the setting of sampling number?
\item \textbf{RQ4:} Whether the performance of CMES is sensitive to the setting of student cluster number?
\item \textbf{RQ5:} How does CMES perform on different ratios of the training set?

\end{itemize}

\subsection{Experimental Settings}

\textbf{Datasets Description. }We conduct experiments on two real-world datasets ASSISTments~\cite{assist} and Math, which both provide student-exercise interaction records and the exercise-knowledge concept relational matrix. ASSISTments is a publicly available dataset collected from the online tutoring system ASSISTments. Math is a proprietary dataset assembled by a renowned e-learning platform, comprising mathematics practice and examination records of elementary and secondary school students. For both datasets, we filter out students with less than $15$ response logs to ensure sufficient data for model learning. After processing, the statistics of the two datasets are shown in Table \ref{dataset_table}. We apply $70\%:10\%:20\%$ training/validation/test split for each student’s response logs in the two datasets.
%assembled by a renowned e-learning platform, comprising mathematics practice and examination records of elementary and secondary school students. 

%我们在两个真实世界的数据集上进行了实验：ASSIST和Math。这两个数据集都提供了学生的交互日志和知识概念。Assist数据集是由ASSISTments在线辅导系统从广泛使用的在线学习系统收集的公开可用数据集。Math数据集是在线教育公司收集的一个私有数据集。我们分别为ASSISTments和MATH过滤掉response logs少于15的学生，以保证每个学生有足够的数据进行诊断。数据集的完整统计信息如表1所示。

\begin{table}[htbp]
    \caption{The statistics of the datasets.} 
    \label{dataset_table}
    \centering
    
  %  \abovecaptionskip 5pt % 设置标题与表格之间的垂直间距
    \begin{tabularx}{1\columnwidth}{@{} l X X @{}}
    %\begin{tabular}{@{} l c c @{}}
        \toprule
        Statistics                    & ASSISTments  & MATH     \\
        \midrule
       \# Students                      & 4,163        & 1,967     \\
       \# Exercises                     & 17,746       & 1,686     \\
       \#  Knowledge concepts            & 123         & 61       \\
        \# Response logs                 & 278,868      & 118,348   \\
        \# Avg logs per student          & 67          & 60            \\
        \bottomrule
    \end{tabularx}
    
\end{table}

\begin{table*}[!ht] 
%\captionsetup[table]{skip=20pt} % 设置大表标题与表格的垂直间距
%\captionsetup[subtable]{skip=1pt} % 设置子表标题与子表的垂直间距
 \centering
%\captionsetup{skip=1pt}
\begin{subtable}{1\linewidth}
\centering

\resizebox{0.95\textwidth}{!}{%
  \begin{tabular}{c|cc|cc|cc}
    \hline
    \multirow{2}{*}{Metrics} & \multicolumn{2}{c|}{ACC} & \multicolumn{2}{c|}{RMSE} & \multicolumn{2}{c}{AUC} \\ \cline{2-7}
    & Orginal-CDMs & CMES-CDMs & Orginal-CDMs & CMES-CDMs & Orginal-CDMs & CMES-CDMs \\ \hline
    IRT &68.89\% &\textbf{70.59}\% &0.4684&\textbf{0.4547}&70.45\% &\textbf{74.40\%} \\
    MIRT & 70.79\% & \textbf{72.36\%} & 0.4634& \textbf{0.4368}& 73.93\% & \textbf{75.55\%} \\
    NCD & 72.27\% & \textbf{72.89\%} & 0.4335& \textbf{0.4283}& 75.22\% & \textbf{76.23\%} \\ 
    CDGK &72.08\% &\textbf{73.01\%} &0.4356&\textbf{0.4306}&74.83\% &\textbf{75.51\%} \\
    ECD & 72.47\% & \textbf{72.80\%} & 0.4334& \textbf{0.4287}& 74.97\% & \textbf{76.25\%} \\
    RCD & 72.99\% & \textbf{73.06\%} & 0.4243& \textbf{0.4237}& 76.40\% & \textbf{76.51\%} \\ \hline
  \end{tabular}%
} \caption*{(a) ASSISTments}
  \label{assist_experiment}
 \end{subtable}%
 
  \vspace{0.2cm} % 添加垂直间距
  
\begin{subtable}{1\linewidth}
\centering

 \resizebox{0.95\textwidth}{!}{%
  \begin{tabular}{c|cc|cc|cc}
    \hline
    \multirow{2}{*}{Metrics} & \multicolumn{2}{c|}{ACC} & \multicolumn{2}{c|}{RMSE} & \multicolumn{2}{c}{AUC} \\ \cline{2-7}
    & Orginal-CDMs & CMES-CDMs & Orginal-CDMs & CMES-CDMs & Orginal-CDMs & CMES-CDMs \\ \hline
    
    IRT &70.88\% &\textbf{72.75\%} &0.4505&\textbf{0.4460}&71.62\% &\textbf{76.37\%} \\
    MIRT & 72.99\% & \textbf{74.60\%} & 0.4284& \textbf{0.4097}& 75.31\% & \textbf{78.04\%} \\
    NCD & 74.13\% & \textbf{74.94\%} & 0.4102& \textbf{0.4053}& 77.14\% & \textbf{78.81\%} \\
    CDGK &73.68\% &\textbf{74.63\%} &0.4121&\textbf{0.4068}&77.00\% &\textbf{78.13\%} \\
    ECD & 74.16\% & \textbf{74.83\%} & 0.4101& \textbf{0.4077}& 77.18\% & \textbf{78.30\%} \\
    RCD & 74.86\% & \textbf{75.16\%} & 0.4063& \textbf{0.4055}& 78.34\% & \textbf{78.64\%} \\ \hline
  \end{tabular}%
}
  \caption*{(b) MATH}
 \label{math_experiment}
\end{subtable}
  \caption{Experimental results on student performance prediction. The best results are highlighted in bold. Our CMES-CDMs significantly outperform the Orginal-CDMs with $p < $0.01.}
   \label{experiment_table}
\end{table*}

\textbf{Evaluation Metrics. }
Considering that there is no true knowledge mastery of students, in the literature, the mainstream approach is to 
indirectly evaluate the effectiveness of CDMs by using the knowledge mastery vector obtained to predict the student’s exercising performance. Three famous metrics, i.e., the Root Mean Square Error (RMSE) \cite{y13}, the Prediction Accuracy (ACC) \cite{y12} and Area Under an ROC Curve (AUC) \cite{auc} were chosen to evaluate predictive performance.
% from the perspectives of regression and classification respectively. From a regression standpoint, RMSE is chosen to quantify the disparity between predicted scores and actual scores. From a classification standpoint, students' exercising performance is grouped into two categories, i.e., correct and incorrect, respectively. 
% We use the Prediction Accuracy (ACC) \cite{ncd} and Area Under an ROC Curve (AUC) \cite{auc} to evaluate the predictive performance of the model.

% to valuate the performance of CDMs from two perspectives: regression and classification. From a regression standpoint, is chosen to quantify the disparity between predicted scores and actual scores. From a classification standpoint, students' correct or incorrect responses are represented as 1 and 0, respectively. We use the Prediction Accuracy (ACC) \cite{ncd} and Area Under an ROC Curve (AUC) \cite{auc} to evaluate the predictive performance of the model.
%主流的CD模型是从回归和分类两个角度采用不同的评价指标来评估 模型的性能。从回归的角度，选择均方根误差（RMSE）来量化预测分数与实际分数之间的差距。而从分类的角度，学生的正确或错误回答可以表示为1和0，我们使用the prediction accuracy (ACC) and area under the curve (AUC) 来评估模型的预测性能。

\textbf{Cognitive Diagnosis Models. }To validate the effectiveness of CMES framework, we conducted the comparison experiments based on six representative CDMs, namely IRT \cite{irt}, MIRT \cite{mirt}, NCD \cite{ncd}, CDGK~\cite{cdgk}, ECD~\cite{ecd} and RCD \cite{rcd}.
%为了验证XXX的有效性，我们在四个经典的认知诊断模型上进行了实验，即IRT、MIRT、NCD和RCD。
 
\textbf{Parameter Settings. }We first initialized all the parameters in the networks with Xavier \cite{xavier} initialization and used the Adam \cite{adam} optimizer with a fixed batch size of $256$ during the training process. For the multi-dimensional models (i.e., MIRT, NeuralCD, CDGK, ECD and RCD), we set the dimensions of latent features for both students and exercises to be equal to the number of knowledge concepts, i.e., $123$ for ASSISTments and $61$ for MATH datasets. 
% The number of sampled exercises $n$ is chosen from $\left \{5, 10, 20, 30, 40\right\}$, and the number of clusters $N$ is chosen from $\left \{20, 30, 40, 50, 60\right\}$. All hyper-parameters of our approach and baselines are tuned on the validation set. 
Based on the parameter tuning, we set $n$ to $20$ for ASSISTments and $5$ for Math respectively; we set $W$ to $50$ and $20$ for ASSISTments and Math respectively. Finally, experimental results for all models are obtained by performing standard $5$-fold cross-validation. The hyper-parameters of comparison approaches are tuned on the validation set according the original paper. All models are implemented in Pytorch, and all experiments are conducted on Linux servers with Tesla V100.
%我们通过PyTorch实现了我们的XXX框架和认知诊断模型，我们首先用Xavier初始化网络中的所有参数，优化器固定为Adam，我们将批量大小设置为256。然后，在多维模型（即MIRT、NeuralCD和RCD）中，我们将学生和习题的潜在特征的维度统一设置为知识概念的数量，即ASSISTments中的123和MATH中的61。所有超参数都在验证集上进行了调整，采样题目的比例$N^{sample}$是从[5,10,20,30,40]中选择的，聚类的簇数$N^{cluster}$是从[20,30,40,50,60]中选择的。基于验证数据集的性能，我们根据ASSISTments和MATH数据集的特点，分别设置两个数据集的参数$N^{sample}_{assist}=20$, $N^{cluster}_{assist}=50$ 和 $N^{sample}_{math}=2$, $N^{cluster}_{math}=20$。最后，通过执行标准的 5 折交叉验证来获得所有模型的实验结果。所有模型都使用 Pytorch 实现，所有实验都在具有Tesla V100的Linux服务器上运行。

\subsection{Performance Comparison (RQ1) }

We compare six pairs of Orginal-CDMs and CMES-CDMs in terms of RMSE, ACC, and AUC. The experimental results are exhibited in Table~\ref{experiment_table}. For each pair, better results are bolded. As shown in the table, for each pair, CMES-CDM outperforms Orginal-CDM in terms of all evaluation metrics on all datasets. Even for RCD that models the intrinsic correlations among knowledge concepts, our CMES can still improve its efficacy. These observations verify that our proposed CMES framework by excavating and leveraging the information within un-interacted exercises can match prevailing CDMs and boost the diagnosis performance of existing CDMs.

%为了展示我们提出框架的性能，我们将不同的认知诊断模型应用到了我们的框架中，实验结果如表1所示，每个CDM有两个子列，左侧显示了未使用我们框架的结果（orginal-CDM），和右侧的使用了CMES框架的结果（CMES-CDM），更好的结果以粗体展示。从表中我们可以看到，对于每个认知诊断模型，在所有数据集上，CMES-CDM在评价指标上都优于orginal-CDM。这表明了我们提出的CMES框架可以匹配主流的CDM，并可以提升模型的性能。即使对于对知识概念内部进行建模的的RCD，我们的CMES也可以对其性能进行优化。总之，我们得出结论，通过挖掘并利用未交互习题中蕴含的信息，CMES可以帮助CDM提升诊断性能。

%\input{AnonymousSubmission/LaTeX/Experiment_tables/vs_random}

\begin{figure}[!t]
    \centering
    \includegraphics[width=1\linewidth]{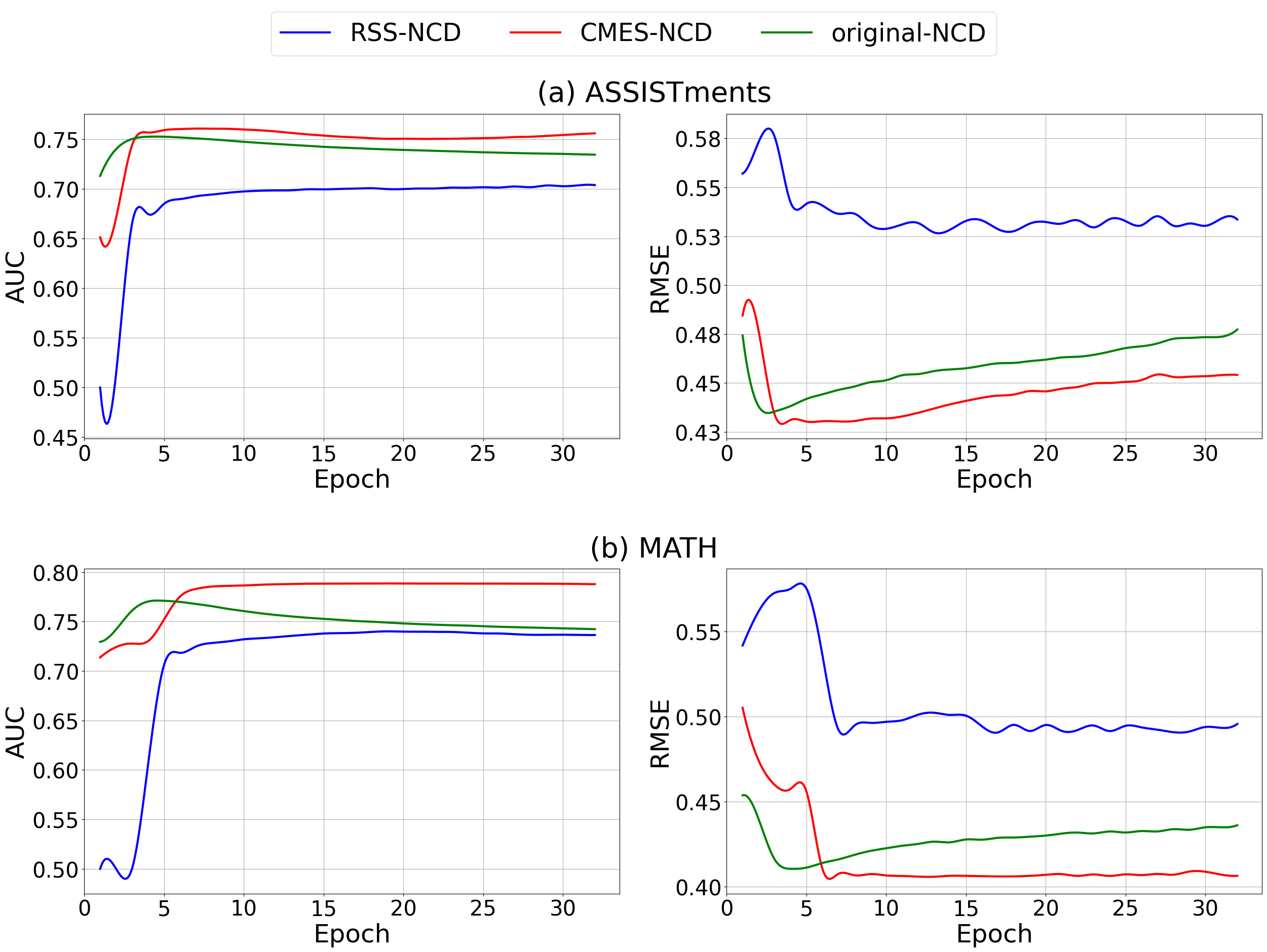}
    \caption{The comparison results between our sampling strategy CMES and the random sampling strategy~(RSS).}
    \label{RQ2}
  
\end{figure}

\subsection{Effectiveness of the Sampling Strategy (RQ2) }

%In CMES, the exercises for diagnosing students are sampled by our sampling module. Instead of directly utilizing the sampled exercises for diagnosis, we fuse the information of sampled exercises to obtain informative ones, which are then used for diagnosis. To validate the efficacy of our module, we design a Random Sampling Module as a variant: it randomly samples exercises from the pool regardless of factors like interacted concepts, and does not fuse information of sampled exercises. The randomly sampled exercises are then input to the Evaluation Module for evaluating potential true labels.
%In XXX, 用于诊断学生的未交互习题来自于我们的采样模块，并不是简单地直接利用这些采样到的习题进行诊断，而是将这些采样到的习题的信息进行融合，从而得到信息丰富的习题再进行诊断。为了证明我们模块的有效性，我们设计了Random Sampling Module as a variant: 我们随机从题库中采样习题，而不考虑是否交互过这些习题关联的知识概念等因素，也不对这些采样的信息进行融合，来代替我们的Sampling Module，然后将随机采样得到的习题输入到Evaluation Module中评估潜在的真实标签。

% In CMES, the instances for diagnosing students originate from our Sample Augmentation Module, rather than simply taking the directly sampled exercises as the training set. 
To validate the effectiveness of the sample augmentation strategy, we compare it with the random sampling strategy~(RSS). RSS randomly samples exercises for each student $s_i$ from $U_{i}$. These sampled exercises are directly used as extra training samples without the information mixture process. The randomly sampled exercises are then fed into the pseudo feedback module to assess the potential labels.
%In CMES, 用于诊断学生的样本来源于我们的Sample Augmentation Module，而并不是简单地直接将采样到的题目作为训练样本。为了证明Sample Augmentation Module的有效性，我们设计了Random Sampling Module as a variant: 我们随机从题库中采样习题，而不考虑是否交互过这些习题关联的知识概念等因素，也不对这些采样的信息进行融合，来代替我们的Sample Augmentation Module，然后将随机采样得到的习题输入到Evaluation Module中评估潜在的真实标签。

Figure~\ref{RQ2} exhibits the comparison results among NCD with our sample strategy~(namely CMES-NCD), NCD with the random sampling strategy~(namely RSS-NCD) and the original NCD~(original-NCD). CMES-NCD markedly surpasses RSS-NCD and original-NCD across all metrics on both datasets, while the random sampling strategy deteriorates model performance.
% , which may be attributed to the excessive redundant exercises attained by random sampling. 
% The comparison manifests that the convergence cycles of SAM are slightly fewer than RSM. 
% Combined with Table~\ref{experiment_table}, despite acquiring pseudo response labels through the Pseudo Feedback Module, RSM deteriorates model performance. This may be attributed to the excessive redundant exercises attained by random sampling. 
The information gathered from these randomly sampled exercises might lack diversity, resulting in ineffective diagnostic values. Even the redundant exercises delude the cognitive diagnosis model, hindering it from accurately diagnosing students' cognitive states. In contrast, the proposed CMES performs collaboration-aware sampling and mixes the information of exercises, enriching diagnostic information to enable the model to infer students' cognitive states more comprehensively.
%表4展示了Sample Augmentation Module(SAM)与Random Sampling Module(RSM)之间的比较。SAM在两个数据集上的所有指标上都明显优于RSM。从比较中我们可以看到，SAM收敛周期略小于RSM。结合表2所示，尽管随机采样得到的样本使用了Pseudo Feedback Module来获得pseudo response labels，但RSM使得模型的性能变差，这可能是因为采样到了过多的冗余习题，且这些题目的信息量单一，无法提供有效的诊断信息，导致这些冗余题目迷惑了认知诊断模型，使得模型无法有效地诊断学生的认知状态。而SAM基于Collaboration-aware采样，又将这些题目的信息进行融合，丰富了诊断信息，使得模型更全面地诊断出学生的认知状态。

\begin{figure}[!t]
    \centering
    \includegraphics[width=1\linewidth]{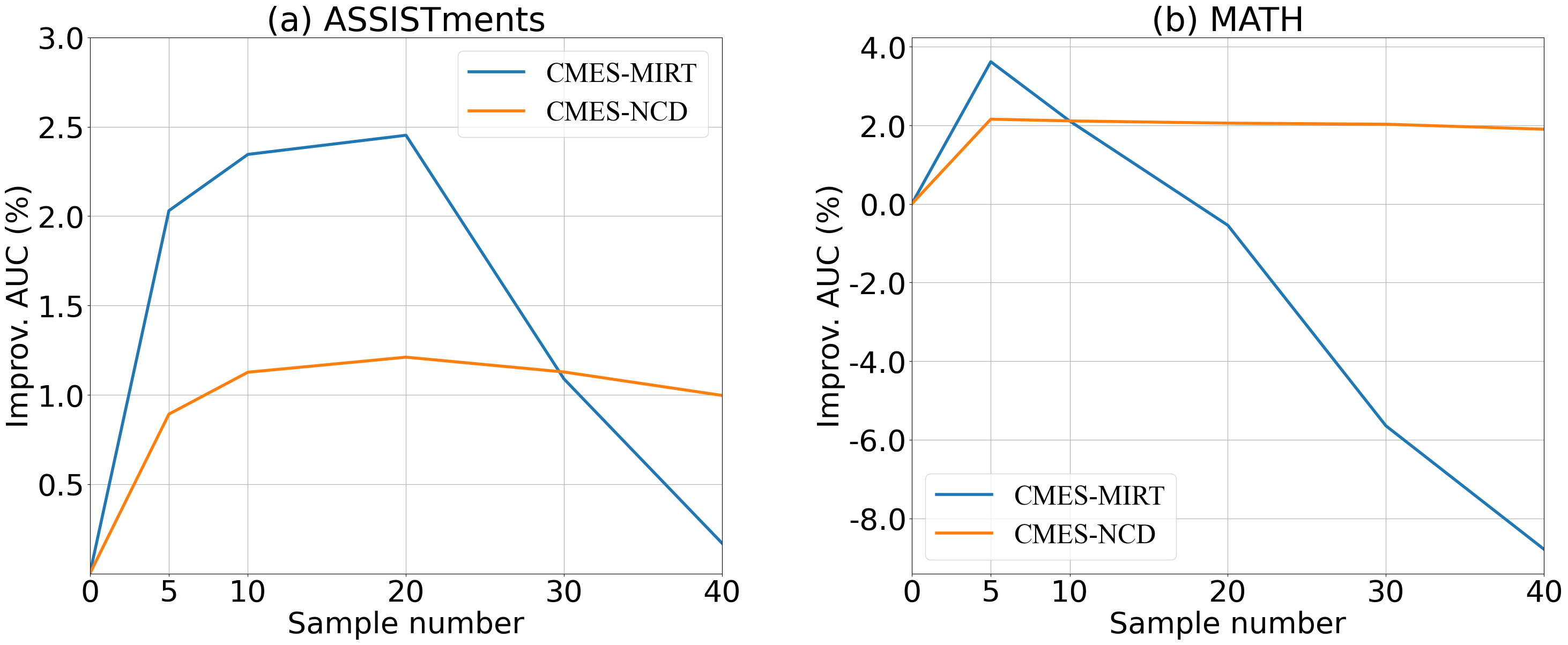}
    \caption{Impact of sampling number.}
    \label{sample_rate_table}
\end{figure}

\subsection{Sensitivity Analysis of Sampling Number (RQ3)}

We chose two representative CDMs~(i.e., MIRT and NCD) combined with our CMES framework to investigate the performance change when varying the sampling number for each student~(i.e., the parameter $n$) in the range of $\left \{ 5,10,20,30,40 \right \}$. As shown in Table \ref{sample_rate_table}, on the ASSISTments, the optimal performance of CMES-MIRT and CMES-NCD is achieved at $n = 20$, while their peak performance is reached at $n=5$ on the Math. It maybe attributed to the exercise pool size of the two datasets. As shown in Table~\ref{dataset_table}, the number of exercise in ASSISTments is more larger than that of Math. The performance of CMES-MIRT and CMES-NCD starts to decrease when $n> 20$ and $n>5$ on ASSISTments and MATH respectively. The degradation is more significant for MIRT, probably because the simple student-exercise interaction function in MIRT model cannot capture fine-grained exercise information. Excessive exercise signals confuse the diagnosis model and lead to negative optimization.

% since sampling more exercises enables richer diagnostic information during mixed. On the MATH, the peak performance of MIRT and NCD models is reached at $i$=5, because users interact with more knowledge concepts on average in MATH, thus fewer sampled exercises can provide informative signals. 
% The performance of MIRT and NCD starts to decrease when $i> 20$ and $i>5$ on ASSISTments and MATH respectively. The degradation is more significant for MIRT, probably because the simpler MIRT model cannot capture fine-grained exercise information. Excessive exercise signals confuse the diagnosis model and lead to negative optimization.

%我们对采样未交互习题个数i的影响进行了实验研究。我们在{5,10,20,30,40}的范围内变化。如表所示，在ASSIST数据集上，MIRT和NCD模型都在i=20时性能达到最好，这是因为采样更多的习题在融合信息时可以更加丰富习题的诊断信息。在MATH数据集上，MIRT和NCD都在i=5时性能达到了最佳，这是因为在MATH数据集中，用户平均交互的知识概念的数量高于ASSIST数据集，因此采样少量的习题可以提供更丰富的信息。MIRT和NCD在两个数据集上的采样个数在分别i>20和i>5时，性能都发生了下降的情况，但是MIRT的性能降低的更明显，这是因为MIRT模型相对于NCD模型更简单，无法捕获到更细腻的习题信息，过多的样本信息混淆了认知诊断模型的诊断能力，因此导致了负优化的结果。

\subsection{Sensitivity Analysis of Student Cluster (RQ4) }

\begin{figure}[]
 \centering
 \includegraphics[width=0.45\textwidth]{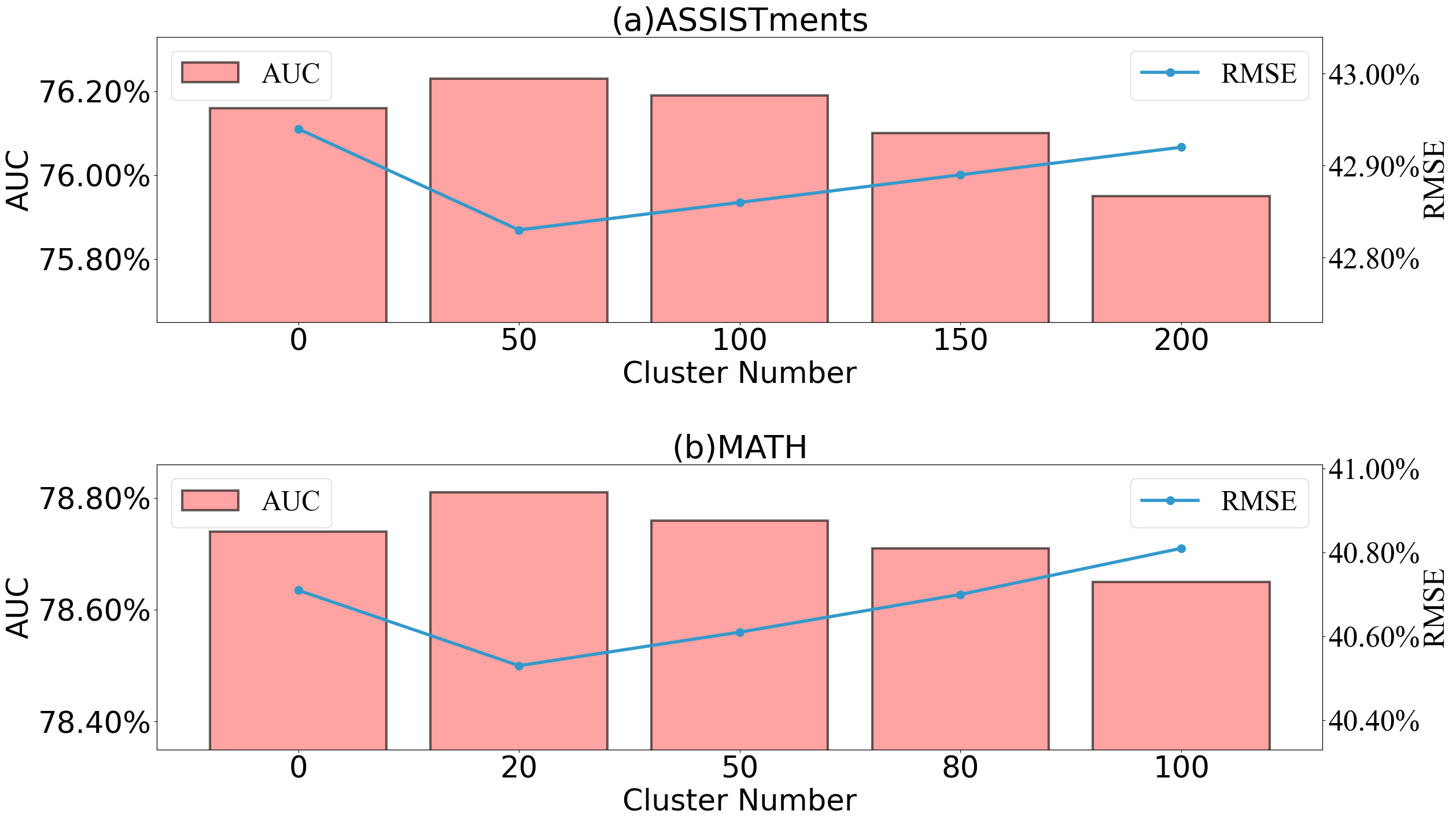}
 \caption{Impact of student cluster number.}
 \label{cluster_number}
\end{figure}

We further used NCD combined with our CMES to probe the impact of the number of clusters $W$. Here we search $W$ in the range of $\left \{ 0,50,100,150,200 \right \} $ and $\left \{0,20,50,80,100 \right \} $ for ASSISTments and MATH respectively. As depicted in Figure~\ref{cluster_number}, the optimal values for $W$ are set to $50$ and $20$ for ASSISTments and MATH. From this observation, on the one hand, the optimal setting for $W$ seems to be related to the student size, as shown in Table~\ref{dataset_table}, the student number in ASSISTments is larger than that in Math. On the other hand, inappropriate setting for the student cluster number $W$ will result in significant performance degradation, which manifests that the performance of CMES is sensitive to the setting of student cluster number and answers RQ4. 
% When $N$ surpasses 50 and 20, the efficacy of CMES in optimizing the model starts declining. 
% This manifests that excessive clusters will divide students of similar proficiency into different groups, thereby expanding the sampling scope and obtaining more exercises with limited information during sampling. Too few clusters tend to lead to redundant exercises. Therefore, both excessive and insufficient numbers of clusters can cause declines in the model's optimization capability to varying extents.
%我们进一步进行了实验来研究聚类数量N的影响。我们使用NCD来进行探索，对于ASSIST和MATH，我们分别搜索N，from $\left \{ 0,50,100,150,200 \right \} $和$\left \{ 0,20,50,80,100 \right \} $。如图~\ref{cluster_number}所示，ASSIST和MATH的最佳值分别是50和20，当时N分别大于50和20时，CMES对模型的优化能力开始下降，这表明过多的簇会导致答题能力相近的学生被分到不同的簇，进而扩大采样范围，在采样时得到了更多的信息量有限的题目，而过少的簇容易得到信息冗余的题目，因此过多和过少的聚类数会使模型的优化性能在不同程度上下降。

\subsection{Case Study (RQ5) }

\begin{figure}[]
 \centering
 \includegraphics[width=0.45\textwidth]{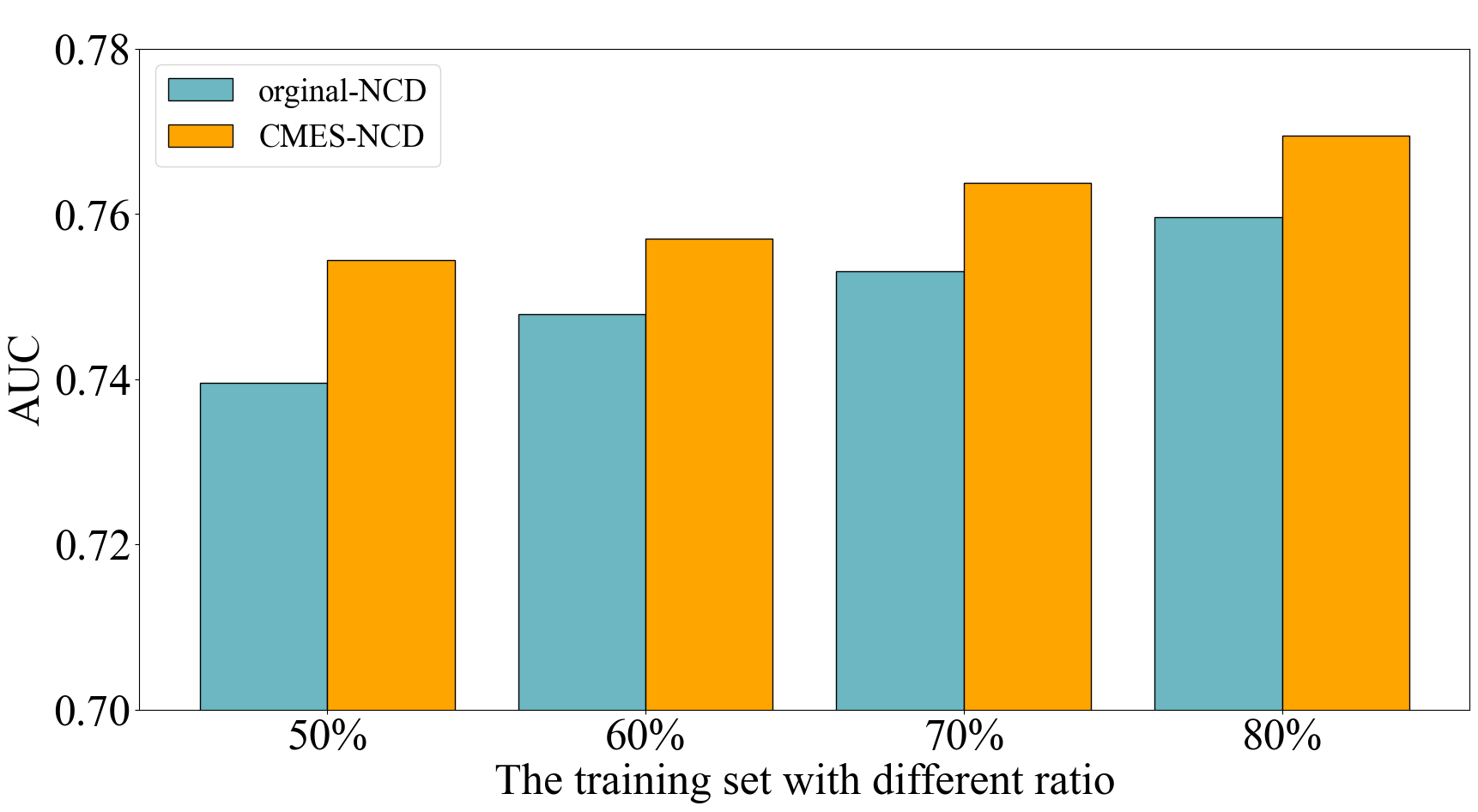}
 \caption{The training set with different ratio.}
 \label{RQ5}
\end{figure}

We selected $20\%$ of the dataset ASSISTments as the test set, and utilized data sets with sizes of $80\%$, $70\%$, $60\%$ and $50\%$ of the full dataset from the remaining data to train model respectively. As depicted in Figure~\ref{RQ5}, we observe that CMES-NCD trained by different size of training sets all demonstrate excellent performance. The performance of CMES-NCD trained on $60\%$ of the data is on par with orginal-NCD trained on $80\%$. Additionally, the enhancement attained by CMES-NCD is most pronounced when trained on $50\%$ of the data, which surpasses the performance of orginal-NCD trained on $70\%$ of the data, and nears its performance trained on $80\%$. These observations validate that our CMES framework can mitigate data scarcity challenges by extracting more information from un-interacted exercises.

%我们将assist数据集划出20%作为测试集，然后用整个数据集的80%、70%、60%和50%的数据(不包含测试集的数据)作为训练集来训练模型。如图~\ref{RQ5}所示，首先我们发现，CMES-NCD在任意比例的训练集上都表现地十分优秀，CMES-NCD在60%的训练集上的表现甚至与orgin-NCD在80%训练集上的表现持平。其次，在50%的训练集上，CMES-NCD提升最为明显，因此证明我们的CMES框架可以通过挖掘更多的习题信息来解决数据稀疏问题，甚至超过了orgin-NCD在70%训练集上的表现，略低于orgin-NCD在80%训练集上的表现。
\section{Conclusion}
In this work, we attempted to explore informative, un-interacted exercises related to broader knowledge concepts with the aim of providing a more comprehensive diagnostic assessment of students. We proposed a generic framework \textbf{CMES}~(\textbf{C}ollaborative-aware \textbf{M}ixed \textbf{E}xercise \textbf{S}ampling) that enables sampling of rich information from un-interacted exercises and facilitates the evaluation of potential true labels. Experimental results on real-world datasets demonstrate the effectiveness of the sampling strategy and the scalability of our framework. We intend to further investigate sampling strategies tailored to the characteristics of cognitive diagnostic models.
%在这项工作中，我们研究了认知诊断任务中的采样策略，旨在更全面地对学生进行诊断。我们设计了一个通用框架，它不仅可以采样到丰富的未交互习题的信息，而且还可以评估潜在的真实标签。在真实数据集上的实验证明了采样策略的有效性和我们框架的可扩展性。我们希望针对认知诊断模型的特点对采样策略进行进一步的研究。
\section{Acknowledgements}
This work was supported in part by the National Natural Science Foundation of China (No. 62107001, No. U21A20512, and No.62302010), in part by the Anhui Provincial
Natural Science Foundation (NO.2108085QF272), in part by the University Synergy Innovation
Program of Anhui Province (NO.GXXT-2021-004), in part by the Key Research and Development Project of Qinghai Province (NO.2023-GX-C13), and in part by the China Postdoctoral Science Foundation (No.2023M740015).
\bibliography{aaai24}

\end{document}